\documentclass[12pt]{article}
\usepackage{amsmath,amssymb}
\renewcommand{\a}{\alpha}
\renewcommand{\b}{\beta}
\newcommand{\bs}{\bigskip}
\newcommand{\dt}{\cdot}

\newcommand{\G}{\Gamma}

\renewcommand{\k}{\kappa}

\renewcommand{\L}{\Lambda}
\newcommand{\mb}{\mbox}

\renewcommand{\th}{\theta}

\begin{document}
\setlength{\baselineskip}{16pt}

\begin{center}
{\large\bf Analysis of Experimental Investigations of}\smallskip\\
{\large\bf Self-Similar Intermediate Structures in}\smallskip\\
{\large\bf Zero-Pressure Boundary Layers at Large Reynolds Numbers}

\vspace{1 truein}

{\sc G. I. Barenblatt,${}^1$ \ A. J. Chorin${}^1$ and V. M.
Prostokishin${}^2$}

\vspace{.33 truein}
${}^1$Department of Mathematics and\\
Lawrence Berkeley National Laboratory \\
University of California\\
Berkeley, California 94720

\bs\bs

${}^2$P. P. Shirshov Institute of Oceanology\\Russian Academy of
Sciences\\
  36, Nakhimov Prospect\\
  Moscow 117218 Russia
\end{center}

\bs\bs\bs\begin{quote}
\begin{center}{\bf Abstract.}\end{center}
Analysis of the Stockholm group data on zero-pressure-gradient
boundary flows, presented in the thesis of J.~M.~\"Osterlund
({\tt http://www.mesh.kth.se/$\sim$jens/zpg/}\,) is performed.
The results of processing of all 70 mean velocity profiles are
presented. It is demonstrated that, properly processed, these data lead
to a conclusion opposite from that of the thesis and related papers:
they confirm the Reynolds-number-dependent scaling law and disprove the
conclusion that the flow in the intermediate (``overlap") region is
Reynolds-number-independent.
\end{quote}
\newpage
\section{Introduction}

Turbulence is the state of vortical fluid flow where velocity,
pressure, and other flow field properties vary in time and space sharply
and irregularly, and can be assumed to be random. The experimental
investigation of  individual realizations of such flows is impossible
because the results are irreproducible:
Experiments repeated under identical external conditions
produce a different outcome. Therefore  experimental investigations of
turbulent flows can only  provide their average properties.

Less trivial and not always recognized is the following: What is of
greatest interest in these experiments  are intermediate asymptotic
states of wider classes of flows, i.e., coherent,
self-consistent fragments common to many different flows. Other
measurements  reflect  special properties of a set-up,  which cannot
 be reproduced in other experiments.
It may sometimes be useful to investigate a particular device
(e.g.~an atomizer) for an immediate
practical purpose, but one should be cautious in
transferring the results to different flows.

Typical examples of intermediate-asymptotic flows are  {\it
shear flows}\/, where the flow is homogeneous in the direction of mean
velocity which depends only on  the coordinate
perpendicular to the mean flow. A well known example of shear flows
occurs in smooth cylindrical circular pipes far from the entrance and
outlet. Another example is the zero-pressure gradient boundary
layer above a smooth flat plate far from its tip. In spite of
their apparent simplicity experiments with such flows require high
experimental culture and are expensive, and therefore relatively rare.
When they are successful, like, for instance, the
experiments of Nikuradze [${}^1$] with flows in smooth pipes in the range
of Reynolds numbers up to
$3.24\dt 10^6$, performed 70 years ago under the guidance of
L.Prandtl, they become milestones in turbulence studies. They
are used to check  theories based on special
hypotheses valid for special classes of flows. Not enough is known
now about the solutions of Navier-Stokes equations to avoid such
hypotheses.

The responsibility of experimentalists who perform such
experiments and process the  data is therefore very high. They are to
be very careful in their conclusions. We showed, for instance
[${}^{2,3}$] that the experiments of Princeton group (A.Smits,
M.V.Zagarola) presented in the thesis of
M.V.Zagarola [${}^4$], which attempted to increase the range of Reynolds
numbers achieved by Nikuradze by an order of magnitude, had a flaw.
Starting from $Re=10^6$, well below the upper boundary of Nikuradze
experiments, their data were influenced by roughness --- insufficient
polishing of the pipe walls.

In the present work we analyze the experiments with
zero-pressure-gradient boundary layers presented in the thesis of
J.M.\"Osterlund [${}^5$]. Like the earlier  theses of
M.V.Zagarola [${}^4$] and M.H.Hites [${}^6$], the thesis of
J.M.\"Osterlund presents the results of  long-time work of a
group headed by a senior scientist (in this case A.V.Johansson),
using a
complicated, expensive and unique facility. The experimental data and
their interpretation presented in this thesis might be accepted by
some readers, as in the case of the thesis by Zagarola, without
 precautions. This was the
motivation of our analysis of this thesis.

The number of runs (70 measurements of mean velocity profiles) reported
in the thesis [${}^5$] is larger than in previously reported series,
although the range of Reynolds numbers covered  is not as wide; less
than in the previous thesis of M.H.Hites [${}^6$]: $2500 < Re_{\theta} <
27,500$. In the thesis [${}^5$] the authors make  very definite
statements: they claim that their experiments confirm the classical
two-layer theory, in particular the Reynolds number-independent universal
logarithmic law, and, exhibiting no Reynolds number-dependence, disagree
with the alternative theory based on the Reynolds number-dependent
scaling law.

J.M.\"Osterlund presented the data of 70  mean velocity measurements
 on the Internet ({\tt www.mesh.kth.se/$\sim$jens/zpg/}\,). We present
here the results of the processing of all these data. We  demonstrate
that, properly processed, these data lead to the opposite
conclusion: they confirm the Reynolds number-dependent scaling law and
disagree with the conclusion of Reynolds number-independence.

\section{Background}

 According to the classical two-layer theory of wall-bounded
turbulent shear flows at large Reynolds numbers, the distribution of
average velocity
$u$ across the flow in the basic intermediate region adjacent to the
viscous sublayer is represented in the form of universal
(Reynolds number-independent) von K\'arm\'an-Prandtl logarithmic law
\begin{equation}\label{1}
\phi =\frac{1}{\k} \ln\eta +C \ .
\end{equation}
Here we use classical Nikuradze-Schlichting et al.~notations:
\begin{equation}
\phi =\frac{u}{u_*} \ , \qquad
u_*=\big( \frac{\tau}{\rho}\big)^{\frac 12} \ , \qquad
\eta =\frac{u_*y}{\nu} \ ,
\end{equation}
where $\tau$ is the shear stress at the wall, $y$ the distance from the
wall; $\nu$ and $\rho$ are the fluid's  kinematic
viscosity and density. In the thesis more modern notations are used:
${\bar U}^+$  instead of
$\phi$, \ $y^+$ instead of $\eta$. So, the
von K\'arm\'an-Prandtl universal law (\ref{1}) is represented in the
thesis in the form
\begin{equation}\label{vonK}
{\bar U}^+=\frac{1}{\k} \ln (y^+)+B \ .
\end{equation}
Here $\k$ (von K\'arm\'an constant) and $B$, according to the logic of
the derivation, should be universal constants identical in all high
quality experiments. It is known from the literature however that
various experiments give substantially different values of these
constants. Nikuradze [${}^1$] determined $\k = 0.417$, \
$B= 5.84$,  Monin and Yaglom [${}^7$] give the values
$\k = 0.40$, \  $B= 5.1$; Schlichting
[${}^8$] gives the values $\k = 0.40$, \  $B= 5.5$. The
difference is substantial, and for many years  doubts have
accumulated on the validity of the universal logarithmic law.

In our papers (see e.g.~[${}^9$],[${}^{10}$]) the derivation of the
universal logarithmic law was reconsidered. It was shown that one of the
basic assumptions is not quite correct, and on the basis of an
alternative assumption, a different ``scaling" (power) law was proposed:
\begin{equation}\label{power}
{\bar U}^+= (C_1\ln Re +C_2)(y^+)^{c/\ln Re }
\end{equation}
where the constants $C_1,C_2,c$ should be universal and
Reynolds number-independent. Comparison with the experimental data of
Nikuradze has given the following values of the constants:
\begin{equation}
C_1=\frac{1}{\sqrt{3}} \ , \qquad
C_2=\frac 52 \ , \qquad  c=\frac 32 \ ,
\end{equation}
so that the law (\ref{power}) is presented in the form
\begin{equation}
\phi = \left( \frac{\sqrt{3}+5\a}{2\a}\right)\eta^\a \ , \qquad
\a=\frac{3}{2\ln Re}
\end{equation}
or, using the notation of \"Osterlund's thesis [${}^5$]
\begin{equation}\label{sqrt}
{\bar U}^+= \left( \frac{1}{\sqrt{3}}\ln Re+\frac 52\right)
(y^+)^{\frac{3}{2 \ln Re}} \ .
\end{equation}
Asymptotically, at $Re\to\infty$, the specific choice of $Re$ is of no
importance: $Re$ can be replaced by $Re'= {\mbox{Const}} \cdot Re$, and
the asymptotic form of (\ref{sqrt}) will remain the same. However, for
practical  large, but not too large, values of $Re$, its
actual expression is significant. It should be remembered that in our
comparison with Nikuradze's data [${}^1$] we used his definition of the
Reynolds number: $Re={\bar u} \ d/\nu$, where ${\bar u}$ is the mean
flow velocity (bulk discharge rate divided by cross-section area), and
$d$ is the diameter of the pipe. Furthermore, it was recognized from the
beginning [${}^{11}$] that the law (\ref{sqrt}) is asymptotic  (in
the parameter $1/\ln Re$). It should be valid at large $Re$, but at
lesser values higher terms in the expansion of the coefficients
could be significant.

The experimental data of the Stockholm group
(J.M.\"Osterlund, A.V.Johansson) are presented in the form of graphs in
the $\ln y^+$, ${\bar U}^+$ plane suggested by the universal logarithmic
law (\ref{vonK}) (see Figure 5 on page 43 and Figures 13, 14 on pages
152--153 of the thesis [${}^5$]). They became
available on the Internet ({\tt www.mesh.kth.se/$\sim$jens/zpg/}\,) in
complete digital form. The data are presented in our table, parameterized
by the authors by the parameter $Re_{\th}=U\th/\nu$. Here $U$ is the free
stream velocity,
$\th$ is the momentum thickness, {\it a quantity measurable a
posteriori}, after the run is performed.

\section{Processing of the mean velocity data}

 The first question to be answered was as follows: are the
mean velocity data  presented
on the Internet compatible with {\it some} scaling law, not
necessarily the law (\ref{sqrt}). Therefore the data were plotted  in the
double-logarithmic coordinates $(\lg y^+,\lg {\bar U}^+)$. The result
was instructive: the data outside the viscous
sublayer form a characteristic shape of a broken line  (see Figures
1--70). This shape is similar to the shape obtained for the experiments
of the first group according to our classification [${}^{12}$] where
there was no influence neither of the external turbulence of the free
stream nor of roughness. The Stockholm authors place the lower
boundary of the intermediate region at $y^+=200$. We found this value
generally too high, and the standard value $y^+=70\div 100$ seems to be
more appropriate, however we marked the line $y^+=200$ on all Figures
1--70.

Thus,  all experiments revealed two straight lines forming a
broken line in the $\lg y^+,\lg U^+$ plane. These straight lines
correspond to the scaling laws
\setcounter{equation}{0}
\begin{equation}\label{I}
\mb{(I)} \quad \overline{U^+} = A(y^+)^\a
\end{equation}
(in the intermediate region adjacent to
the viscous sublayer), and
\begin{equation}\label{II}
\mb{(II)} \quad \overline{U^+} = B(y^+)^\b
\end{equation}
(in the intermediate region adjacent to
the free stream). The coefficients $A,\a,B,\b$ were obtained by
standard statistical processing of data (see the table). The
coefficients $A,\a$ of the straight line (\ref{I}) representing the
scaling law for the mean velocity distribution in the basic intermediate
region adjacent to the viscous sublayer are obviously
Reynolds-number-dependent: For us this was not unexpected, because
previous processing of all available experimental data  for a
much wider range of Reynolds number led us to the same conclusion
(see paper [${}^{12}$] which was known to the Stockholm group
and referenced by them). Therefore we  conclude that the
validity of {\it some} Reynolds-number-dependent scaling law for
 mean velocity distribution is unquestionably confirmed  by
the experiments of the Stockholm group as well.

Note that because the Reynolds number range covered by the Stockholm
group was not large, substantially less than  the range covered by the
other groups, in particular the Chicago group [${}^6$], there would be a
danger that they would not  notice the
 Reynolds-number-dependence because the governing parameter is $\ln Re$,
not $Re$ itself. This is not the case: The $Re$-dependence of the
Stockholm experimental data is sufficiently strong to be revealed by
proper processing.

By the way, the authors could notice that their values $\k=0.38$ and
$\b =4.1$ are substantially less than those presented in the literature
as standard. However, if the law is universal
Reynolds-number-independent, these parameters should be identical for
all experiments of sufficient quality!

The argument against the power law used by the authors (see paper
[${}^{13}$] reproduced in the thesis) is the following.
They introduce the ``diagnostic function"
\[
\G =\frac{y^+}{{\bar U}^+} \ \frac{d{\bar U}^+}{dy^+} \ .
\]
Their statement, ``The function $\G$ should be a constant in a region
governed by a power law" is correct for a fixed Reynolds number.
However, this is not true for the ``diagnostic function  averaged for KTH
data", which is shown in their Figure 6.

We invite the reader to look at any of the Figures 1--70.
It is clear that for each run $\G$ is a constant --- look at  the
straight lines in the first intermediate region! However,
{\it this constant is different for different runs because the slopes of
straight lines is $Re$-dependent!} Indeed, the slope in the first region
decays with growing  Reynolds number. It is clear why $\G$ obtained by
the authors is decreasing: the runs with larger Reynolds number and
smaller slopes contribute more to larger $y^+$.

 Now, when the validity of  {\it some}
Reynolds-number-dependent scaling law for the experiments of
the Stockholm group is unquestionably established, we have to
investigate whether this scaling law can be represented in the same
form (\ref{sqrt}) obtained for  flows in pipes. But what is Re? We
cannot assume it arbitrarily  to be equal to $Re_{\th}$.

This effective Reynolds number $Re$ should have the form
$Re=U\L/\nu$, where $U$ is the free stream velocity, $\nu$ the
kinematic viscosity, and $\L$ a  length scale which we cannot
{\it \`a priori} identify with the momentum thickness $\th$, as there is
no rationale for such identification. So, the basic question is whether
one can find for each run such length scale $\L$ so that the scaling law
(\ref{sqrt}) will be valid for the mean velocity  distribution in the
first intermediate region. A priori the very existence of such a length
scale is under question, but if it does exist, this means that the law
(\ref{sqrt}) is not  specific to flows in pipes but can be a
general law for wall-bounded shear flows at large Reynolds numbers.

To answer this question one has to take the values $A$ and $\a$ for
each run, obtained by standard statistical processing of the
experimental data in the first intermediate scaling region, and then
calculate two values $\ln Re_1$ and $\ln Re_2$ by solving two equations
suggested by the scaling law (\ref{sqrt}),
\begin{equation}\label{two}
\frac{1}{\sqrt{3}} \ln Re_1 +\frac 52 = A \ , \qquad
\frac{3}{2\ln Re_2}=\a \ .
\end{equation}
If the values $\ln Re_1$ and $\ln Re_2$  obtained by solving
two different equations (\ref{two}) are close, i.e.~if they coincide
within experimental accuracy, it will mean that the unique length scale
$\L$ can be determined so that the experimental scaling law in the
region (\ref{I}), whose existence was proved before, coincides with the
basic scaling law (\ref{sqrt}). The table shows that indeed these
values  are close --- for all $Re_{\th}>10,000$, the difference
$\ln Re_1-\ln Re_2$ does not exceed 3\%. This allows one to introduce
for large Reynolds numbers the effective Reynolds number $Re$ according
to the relation
\begin{equation}\label{geom}
\ln Re_1 ={\textstyle{\frac 12}} (\ln Re_1+\ln Re_2) , \quad {\mb{or}}
\quad Re=\sqrt{Re_1 Re_2} \ ,
\end{equation}
i.e., the geometric mean of $Re_1$ and $Re_2$. This Reynolds number
allows the definition of  the effective length scale $\L$, which plays a
similar role in the scaling law for the boundary layer flow as does the
pipe diameter for flow in pipes. We remind the reader that the momentum
thickness is calculated by integration of the  velocity
profile obtained experimentally: the  calculation of the
length scale on the basis of the measured velocity profile is not more
complicated. Naturally the ratio of  two length scales $\th/\L$ is
different for different runs: both these quantities depend upon the
details of flows, in particular they can depend in principle upon the
distance between the tip of the plate and the point of observation.

\section{Checking universality}

The scaling law (\ref{sqrt}) can be reduced to a universal form
\setcounter{equation}{0}
\begin{equation}\label{univ}
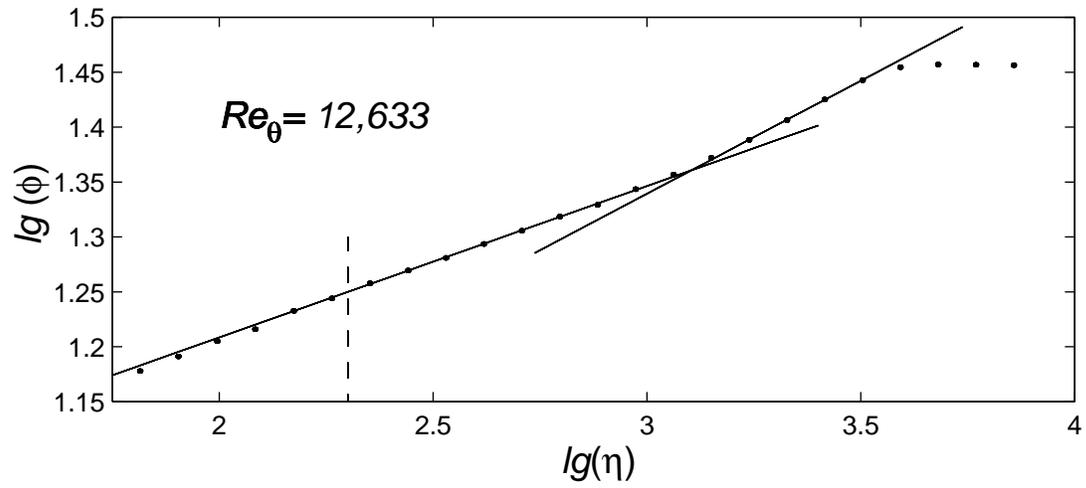
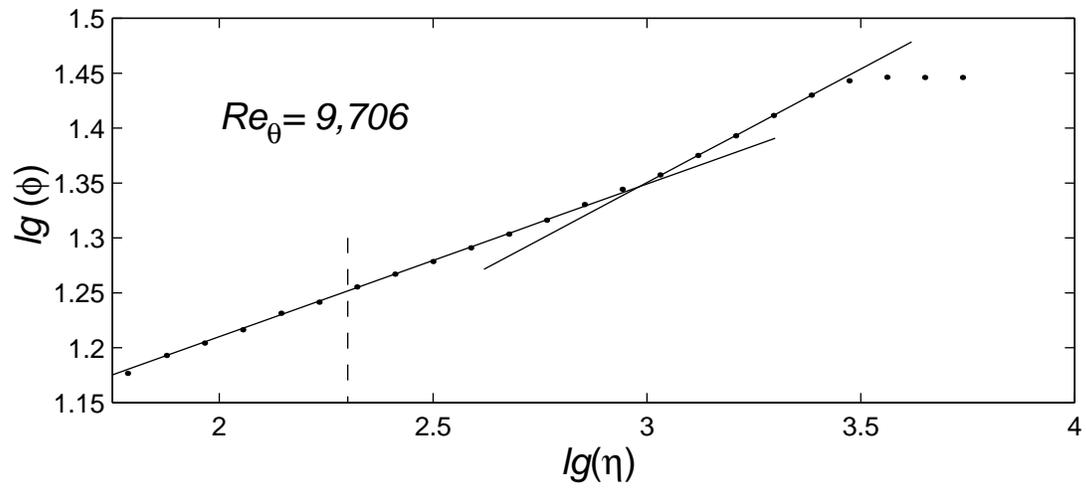
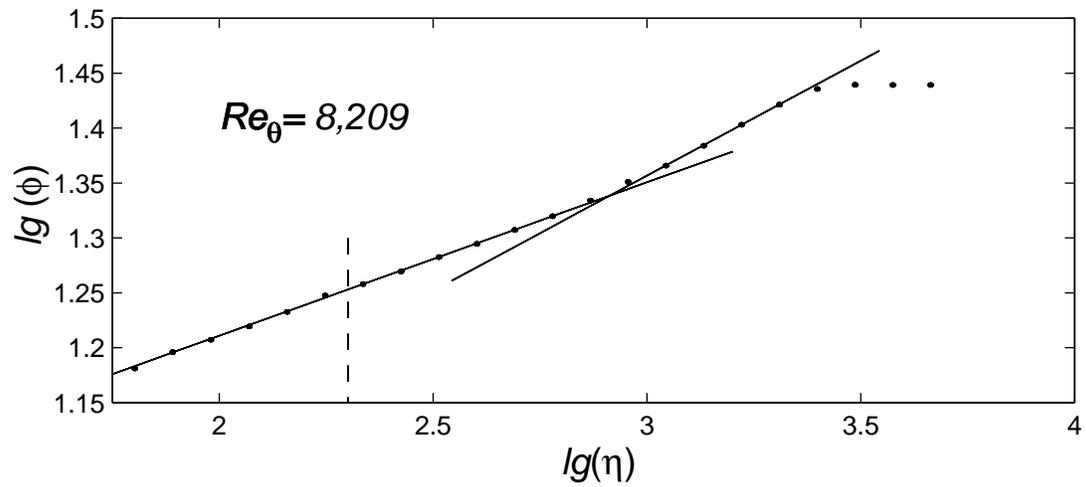
\psi =\frac{1}{\a}\ln\left( \frac{2\a {\bar U}^+}{\sqrt{3}+5\a}
\right)=\ln y^+
\end{equation}
where $\a=\frac{3}{2 \ln Re}$. This formula gives another way to check
the applicability of the Reynolds-number-dependent scaling law
(\ref{sqrt}) in the intermediate region (\ref{I}). Indeed, according to
(\ref{univ}), in the coordinates $\ln y^+,\psi$, all experimental points
should collapse onto the bisectrix of the first quadrant. Figure 71
shows that all data for large Reynolds numbers ($Re_{\th}>15,000$, \
24 runs) presented on the Internet collapse onto the bisectrix with
accuracy sufficient to give an additional confirmation to the
 Reynolds-number-dependent scaling law (\ref{sqrt}).
For lesser values of $Re_{\th}$ a systematic parallel shift is observed
(Figures 72--74). Apparently in this case the choice of $Re$ according
to  (\ref{geom}) is insufficient because at small Reynolds numbers the
higher terms of the expansion could have some influence (see the paper
by Radhakrishnan Srinivasan [${}^{13}$]).

\section{Conclusion}

 The thesis of J.\"Osterlund contains the following statements:
\begin{enumerate}
\item (p.22 of thesis), ``The classical two layer theory was confirmed
and constant values of the slope of the logarithmic overlap region
(i.e.~von K\'arm\'an constant) and additive constants were found and
estimated to  $\k =0.38$, \ $B=4.1$, and $B_1=3.6$."
\item (p.29 of thesis), in fact the Introduction to the paper [14]:
``Contrary to the conclusions of some earlier publications, careful
analysis of the data reveals no significant Reynolds number dependence
for the parameters describing the overlap region using the classical
logarithmic relation."
\end{enumerate}

These statements are not correct. Our careful analysis of experimental
data presented in the thesis performed in the present work leads to the
opposite conclusions.
\begin{quote}
$1'.$ The results contradict the classical two-layer theory.
The estimates of the constants obtained by the authors are substantially
different from the standard values and  this reason alone is enough to
reject the assumption of universality, the cornerstone of the classical
theory.

\medskip
$2'.$ In full agreement with our earlier publications, careful
analysis of the data reveals significant Reynolds number dependence for
the parameters describing the ``overlap" region and confirms the
Reynolds-number-dependent scaling law.
\end{quote}

 The thesis of J.\"Osterlund was not the first investigation
of this kind, many similar experimental investigations were performed
earlier covering a much larger range of Reynolds number. (In the thesis
only one decade of $Re_{\th}$ was covered: $2,500 < Re_{\th} < 27,000$;
in previous investigations, in particular in those reflected in the
instructive review [${}^{15}$], the range $1,000 < Re_{\th} < 200,000$
was covered). However,  as we showed above, the accuracy of
experimental data is sufficient to reveal Reynolds number dependence and
correspondence of their data to the Reynolds-number-dependent scaling
law (\ref{sqrt}) proposed by us.

 The experiments of \"Osterlund et al.~allow an additional
confirmation of the separation of the basic part of the flow into two
self-similar regions (I) and (II) governed by the laws (\ref{I}) and
(\ref{II}). It is important that these two self-similar regions cover
the whole boundary layer and not a small (1/6!) part of it where the
universal law is expected by the authors to be valid.
 These experiments  reveal  a weak
$Re$-number dependence of the parameter $\b$ (Figure 75): it decreased
with growing $Re$. The data are not sufficient to come to a final
decision, but they are in approximate agreement with the correlation
\setcounter{equation}{0}
\begin{equation}
\b=\frac{2}{\ln Re} + 0.01 \ .
\end{equation}

\bs{\bf Acknowledgments.} This work was supported in part by the
Applied Mathematics subprogram of the U.S.~Department of Energy under
contract DE--AC03--76--SF00098, and in part by the National Science
Foudnation under grants DMS\,94--16431 and DMS\,97--32710.

\newpage\begin{center}{\Large\bf References}\end{center}

\begin{enumerate}\item
Nikuradze, J. 1932. Zur turbulenten Str\"omung in glatten Rohren. VDI
Forschungsheft, no.~356.

\item Barenblatt, G.I., Chorin, A.J., and Prostokishin, V.M. 1997.
Scaling laws in fully developed turbulent pipe flow: discussion of
experimental data. {\it Proceedings National Academy of Sciences USA}
{\bf 94}a, pp.773--776.

\item Barenblatt, G.I., Chorin, A.J. 1998. Scaling of the
intermediate region of wall-bounded turbulence: The power law.
{\it Physics of Fluids} \ vol.~{\bf 10}, no.~4, pp.1043--1044.

\item Zagarola, M.V. 1996. Mean-flow scaling of turbulent
pipe flow. Doctoral Thesis, Princeton University, Princeton, New Jersey.

\item
\"Osterlund, J.M. 1999. Experimental studies of zero pressure-gradient
turbulent boundary layer flow. Doctoral Thesis, Royal Institute of
Technology, Stockholm.

\item
Nagib, H. 1997. Scaling of high Reynolds number turbulent boundary
layers in the National Diagnostic Facility. Doctoral Thesis, Illinois
Institute of Technology, Chicago.

\item Monin, A.S. and Yaglom, A.M. 1971. {\it Statistical Fluid
Mechanics}, vol.~1, MIT Press, Boston.

\item Schlichting, H. 1968. {\it Boundary Layer Theory},
McGraw-Hill, New York.

\item Barenblatt, G.I., Chorin, A.J., and Prostokishin, V.M. 1997.
Scaling laws in fully developed turbulent pipe flow. {\it Applied
Mechanics Reviews} \ vol.~{\bf 50}, no.~7, pp.413--429.

\item Chorin. A.J. 1998. New perspectives in turbulence.
{\it Quarterly of Applied Mathematics}, vol.~{\bf LVI}, no.~4,
pp.767--785.

\item Barenblatt, G.I., 1993. Scaling laws for fully developed
turbulent shear flows. Part 1: Basic hypotheses and analysis.
{\it Journal of Fluid Mechanics}, vol.~{\bf 248}, pp.513--520.

\item  Barenblatt, G.I., Chorin, A.J., and Prostokishin, V.M. 1999.
Self-similar intermediate structures in turbulent boundary layers at
large Reynolds numbers. UC Berkeley CPAM preprint 775. {\it Journal of
Fluid Mechanics, in press.}

\item Srinivasan, Radhakrishnan, 1998. The importance of higher-order
effects in the Barenblatt-Chorin theory of wall-bounded fully developed
turbulent shear flows.
{\it Physics of Fluids} \ vol.~{\bf 10}, no.~4, pp.1037--1039.

\item \"Osterlund, J.M., Johansson, A.V.,  Nagib, H.M. and Hites, M.H.
2000. A note on the overlap region in
turbulent boundary layers. {\it Physics of Fluids}, vol.~{\bf 12},
no.~1, pp.1--4.

\item
Fernholz, H. H. and Finley, P. J., 1996. The incompressible
zero-pressure-gradient turbulent boundary layer: an assessment of the data.
{\it Progr. Aerospace Sci.} {\bf 32}, pp.245--311.
\end{enumerate}

\end{document}